\documentclass[rmp,aps,twocolumn,floatfix,authordate1-4]{revtex4}

\usepackage{times}
\input psfig.sty
\usepackage{graphicx}  

\begin{document}
\draft

\title{Cationic DMPC\,/\,DMTAP Lipid Bilayers: Molecular Dynamics Study} 

\setcounter{page}{1}

\author{A.\,A. Gurtovenko}
\affiliation{Laboratory of Physics and 
Helsinki Institute of Physics, Helsinki University of Technology,
P.\,O. Box 1100, FIN--02015 HUT, Finland, and 
Institute of Macromolecular Compounds, Russian Academy of Sciences,
Bolshoi Prospect 31, V.\,O., St. Petersburg, 199004 Russia
}

\author{Michael Patra}
\author{Mikko Karttunen}
\affiliation{Biophysics and Statistical Mechanics Group,
Laboratory of Computational Engineering, Helsinki University
of Technology, P.\,O. Box 9203, FIN--02015 HUT, Finland}

\author{Ilpo Vattulainen}
\affiliation{Laboratory of Physics and Helsinki Institute of Physics,
Helsinki University of Technology, P.\,O. Box 1100, FIN--02015 HUT, Finland}

\date{December 16, 2003}

\begin{abstract}
Cationic lipid membranes are known to form compact complexes 
with DNA and to be effective as gene delivery agents both in 
vitro and in vivo. Here we employ molecular dynamics simulations 
for a detailed atomistic study of lipid bilayers consisting of 
a mixture of cationic dimyristoyltrimethylammonium propane 
(DMTAP) and zwitterionic dimyristoylphosphatidylcholine (DMPC). 
Our main objective is to examine how the composition of the 
bilayers affects their structural and electrostatic properties 
in the liquid-crystalline phase. By varying the mole fraction of 
DMTAP, we have found that the area per lipid has a pronounced 
non-monotonic dependence on the DMTAP concentration, with 
a minimum around the point of equimolar mixture. We show that 
this behavior has an electrostatic origin and is driven by the 
interplay between positively charged TAP headgroups and the 
zwitterionic PC heads. This interplay leads to considerable 
re-orientation of PC headgroups for an increasing DMTAP 
concentration, and gives rise to major changes in the 
electrostatic properties of the lipid bilayer, including 
a significant increase of total dipole potential across the 
bilayer and prominent changes in the ordering of water in the 
vicinity of the membrane. Moreover, chloride counter-ions are 
bound mostly to PC nitrogens implying stronger screening of PC 
heads by Cl ions compared to TAP head groups. The implications 
of these findings are briefly discussed. 
\end{abstract}

\maketitle

\section{Introduction}

Gene therapy based on the introduction of genetic material 
into cells is one of the most promising biomedical approaches 
to treat human diseases \cite{RLanger01,Monkkonen:1998,Gennes:1999}. 
The majority of the delivery vectors proposed are of viral nature. 
The viral vectors have been demonstrated to be very efficient 
but their use is restricted by accompanied toxicity \cite{And98}. 
This has stimulated a search for non-viral delivery systems 
which should be characterized by greater safety and ease of 
manufacturing \cite{RLanger01}. Numerous examples of non-viral 
delivery vectors include cationic liposomes, cationic polymers 
(such as polyamidoamine dendrimers, polyethylenimine, and spermine), 
and block copolymers 
\cite{Pit97,Monkkonen:1998,Sme00,Kuk96,Ast96,Gao96,Fis99}.

In the light of the above, it is surprising how little attention 
has been devoted to computational studies of membranes containing 
cationic lipids. Bandyopadhyay et al.~\cite{Ban99} performed an 
atomistic molecular dynamics (MD) study of a mixture of 
dimyristoylphosphatidylcholine (DMPC) and 
dimyristoyltrimethylammonium propane (DMTAP) 
in the presence of a short DNA fragment. Apart from the 
very elegant piece of work above there are, to the best of our 
knowledge, no published atomistic computational studies of systems 
containing cationic lipids -- this is very much in contrast to the 
great number of computational studies of various neutral and anionic 
phospholipid bilayer systems \cite{Fel00,Sai02a,Sai02b,Tie97,Tob01}. 
Another related example is the recent molecular dynamics study of
B\"ockmann et al.~\cite{Boeckmann:2003} who showed the importance 
of monovalent ions on the properties and organization of lipid 
membranes -- ions are always present in cationic lipid systems. 
The above examples demonstrate that detailed molecular dynamics 
studies can provide valuable insight into the atomistic organization 
of systems containing cationic lipids and yield useful information 
for experimentalists about the underlying mechanisms on the atomic 
and molecular levels.

\begin{figure}[tb]
\includegraphics[width=3.0cm,bbllx=0,bblly=0,bburx=410,bbury=605,clip=]
{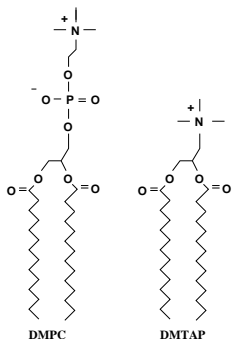}
\caption{\label{fig.1}
Chemical structures of the two lipids 
considered in the present work: a zwitterionic
dimyristoylphosphatidylcholine (DMPC) and a cationic
dimyristoyltrimethylammonium propane (DMTAP).}
\end{figure}

In the present work, our objective is to gain insight into the 
structural and electrostatic properties of cationic lipid bilayers 
through atomic classical molecular dynamics simulations. We 
concentrate on a bilayer mixture composed of two kinds of 
lipids: Neutral (zwitterionic) DMPC and cationic DMTAP 
(see Fig.~\ref{fig.1} for their chemical structures). Since DMTAP 
is positively charged under physiological conditions, we have 
neutralized its positive charges by chloride counter-ions. From 
the computational point of view, this choice for a model system 
is motivated by the fact that DMPC and DMTAP have the same 
non-polar hydrocarbon chains and differ only by their headgroups. 
On the practical side, DMPC\,/\,DMTAP binary lipid mixtures have 
been widely studied in the presence of DNA by various experimental 
techniques~\cite{Art98,Zan99a,Zan99b,Poh00}, and also by 
computational methods~\cite{Ban99}.

We focus mostly on how the composition of the cationic lipid 
bilayer affects the structural and electrostatic properties of 
these lipid bilayer systems. To this end, we consider mixtures 
of DMPC and DMTAP with various different mole fractions of the 
cationic DMTAP component under conditions corresponding to the 
liquid-crystalline phase. We find that DMTAP plays a prominent 
role leading to considerable changes compared to the pure DMPC 
bilayer. As discussed in the present study, this is characterized 
by the strong interplay between electrostatics and structural 
changes in the vicinity of the membrane-water interface. In 
particular, we find that DMTAP gives rise to a non-monotonic 
dependence of the average area per lipid on the DMTAP concentration, 
substantial changes in the electrostatic profile of the membrane, 
and significant re-orientation of P\,--\,N dipole vectors in DMPC 
headgroups. The spatial re-arrangement of P\,--\,N dipoles is 
particularly interesting as it likely plays a significant role 
in the stability of DNA-membrane complexes.

\section{System}

\subsection{Model and simulation details}

We have performed atomistic simulations of fully hydrated 
lipid bilayers consisting of a mixture of cationic DMTAP and 
zwitterionic DMPC lipids. In all simulations, the total number 
of lipid molecules was fixed to 128, and the number of water 
molecules ranged from 3527 (pure DMTAP) to 3655 (pure DMPC).

Force-field parameters for the lipids were taken from the 
recent united atom force-field~\cite{Berger97}. This force-field 
has been previously validated~\cite{Tiel96,Lin00} and is 
essentially based on the GROMOS forcefield for lipid head groups, 
the Ryckaert-Bellman's potential~\cite{Ryck75,Ryck78} 
for hydrocarbon chains, and the OPLS parameters~\cite{Jorg88} 
for the Lennard-Jones interactions between united CH$_n$ groups 
of acyl chains re-parameterized for long hydrocarbon chains 
to reproduce experimentally observed values of volume per 
lipid~\cite{Nag88}. The parameters for this force field are 
available on-line at 
http://moose.bio.ucalgary.ca/Downloads/files/lipid.itp. 
Water was modeled using the SPC water model~\cite{Ber81}. 
The unit positive charge carried by each DMTAP molecule is 
compensated by the introduction of the corresponding number 
of explicit Cl$^-$ counter-ions. While being aware of the 
effects of different models for chloride~\cite{Patra04}, we 
decided to use the default set of chloride parameters 
supplied within the Gromacs force field~\cite{Ber95,Lin01}.

Following the original parameterization~\cite{Berger97}, 
the Lennard-Jones interactions were cut off at 1\,nm 
without shift or switch function. Since long-range electrostatic 
interactions are essential in the present study, and since 
truncation of these interactions has been shown to lead to 
artifacts in simulations of phospholipids bilayers~\cite{Patra03}, 
we employ the particle-mesh Ewald (PME) method~\cite{Dar93}. 
The long-range contribution to the electrostatics is updated 
every 10-th time step.

The simulations were performed in the $NpT$ 
ensemble. The temperature was kept constant using 
a Berendsen thermostat~\cite{Ber84} with a coupling time 
constant of 0.1\,ps. Lipid molecules and water (including 
counter-ions) were separately coupled to a heat bath. Pressure 
was controlled by a Berendsen barostat~\cite{Ber84} with a coupling 
time constant of 1.0\,ps. Pressure coupling was applied 
semi-isotropically: The extension of the simulation box in 
the $z$ direction  (i.\,e., in the direction of the bilayer 
normal) and the cross-sectional area of the box in the $x$-$y$ 
plane were allowed to vary independently of each other.

We considered 11 DMPC\,/\,DMTAP mixtures ranging from pure 
DMPC to pure DMTAP. The molar fractions of the cationic DMTAP, 
$\chi_{\text TAP}$, were taken to be 0.0, 0.06, 0.16, 0.25, 0.31, 
0.39, 0.50, 0.63, 0.75, 0.89, and 1.0.

The main transition temperature of a pure DMPC bilayer is 
$T_m = 24\,^{\circ}$C~\cite{Cevc87}. For DMPC\,/\,DMTAP binary 
mixtures it has been found~\cite{Zan99b} that the main 
transition temperature changes non-monotonically with the mole 
fraction of DMTAP, demonstrating a maximum of approximately 
$37\,^{\circ}$C at $\chi_\mathrm{TAP} \simeq 0.45$. All our simulations 
were done at a temperature of $50\,^{\circ}$C, such that the 
bilayers are in the liquid-crystalline phase.

All bond lengths of the lipid molecules were constrained to 
their equilibrium values using the LINCS algorithm~\cite{Hes97} 
whereas the SETTLE algorithm~\cite{Miy92} was used for water 
molecules. The time step in all simulations was set to 2\,fs.
All simulations were performed using the GROMACS 
package~\cite{Ber95,Lin01}. The combined simulated time of 
all simulations amounts to 250\,nanoseconds. Each simulation 
was run in parallel over 4\,processors on an IBM eServer Cluster 
1600 system. In total, the simulations took about 20,000 hours 
of CPU time.

\subsection{Simulation setup}

Mixtures of DMPC and DMTAP were prepared and equilibrated 
in several steps as follows:

\begin{enumerate}

\item  
We used the equilibrated DPPC bilayer structure of Patra 
et al.~\cite{Patra03} as our initial configuration. The 
structure is available electronically at 
http://www.softsimu.org/downloads.shtml.

\item  
DPPC and DMPC molecules differ only by length of their tail.  
Thus, we created a DMPC bilayer by shortening the DPPC acyl 
chains by two hydrocarbons. This procedure does not disturb 
the acyl chain region or the water-lipid interface.

\item  
The next step was to compress the DMPC bilayer in order to eliminate 
the gap between leaflets created in the previous step. For 
this purpose a pre-equilibration run for 1\,ns in the $NpT$ 
ensemble was performed. After this, the gap between leaflets 
had disappeared, and the obtained DMPC bilayer structure was 
used as initial configuration for all DMPC\,/\,DMTAP mixtures 
described in the following.

\item  
The chemical structures of DMPC and DMTAP differ only by their 
head groups, see Fig.~\ref{fig.1}. With that in mind, the following 
procedure was used to prepare mixtures. For each DMTAP concentration 
the corresponding number of randomly chosen PC headgroups in a pure 
DMPC bilayer were converted to TAP head groups~\cite{Ban99}. To 
neutralize the unit charges in DMTAP headgroups, randomly chosen 
water molecules were replaced by chloride ions while ensuring 
a minimum separation of 0.5\,nm between ions. To retain symmetry 
between the two leaflets, each of them contains the same number 
of cationic DMTAP. 

\item  
Since TAP headgroups occupy a smaller volume than those of DMPCs, 
a short 10\,ps run in the $NVT$ ensemble was performed to let the 
water molecules adjust at the lipid-water interface. This step 
completes pre-equilibration. 

\item  
The actual equilibration runs were performed in the $NpT$ 
ensemble. The needed equilibration times prior to actual production 
runs ranged from 10\,ns to 20\,ns depending on the DMTAP mole 
fraction. We concluded that equilibration was completed when the 
average area per lipid had become stable and fluctuated around 
its mean with a standard deviation not exceeding the standard 
deviation for a pure DMPC bilayer. 
\end{enumerate}

After equilibration, for each DMPC\,/\,DMTAP mixture, 
we performed a production run of 10\,ns in the $NpT$ ensemble 
to collect the data for analysis.

\section{Results and Discussion}

\subsection{Area per lipid}

The average area per lipid, $\langle A\, \rangle$, is one 
of the most fundamental characteristics of lipid bilayers~\cite{Nag00}. 
While being one of the rather few structural quantities that can be 
measured accurately from model membranes via experiments, its also 
plays a major role in a number of quantities, including the ordering 
of acyl chains and the dynamics of lipids in a bilayer. Further, from 
computational point of view, it is highly useful as a means of 
monitoring the equilibration process.

\begin{figure}[tb]
\vspace*{-1.6cm}
\includegraphics[width=6.5cm,clip=true,viewport=0 0 80 100]
{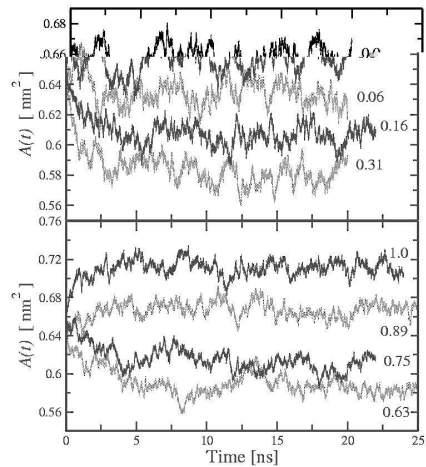}
\caption{\label{fig.2}
Time evolution of the area per lipid, $A(t)$, for different
mixtures of DMPC and DMTAP. The low concentration results 
($\chi_\mathrm{TAP} \leq 0.31$) are shown on the top, and the high 
concentration regime ($\chi_\mathrm{TAP} \geq 0.63$) is illustrated 
at the bottom. For clarity's sake, the results at intermediate 
concentrations are not shown here.}
\end{figure}

Due to the lack of experimental data for the average area per 
lipid in binary DMPC\,/\,DMTAP bilayer mixtures, or in a pure 
DMTAP bilayer [apart from the low-temperature phase \cite{Lew01}], 
reproduction of the experimental data available for pure DMPC 
membranes is essential to validate our approach. To this end, 
let us first consider the temporal behavior of the area per 
lipid, $A(t)$, presented in Fig.~\ref{fig.2}. It shows that 
the obtained average area per lipid for a pure DMPC system has 
a value of $\langle A\, \rangle = 0.656\pm 0.008$\,nm$^2$, in 
very good agreement with the experimentally observed value of 
0.654\,nm$^2$ \cite{Pet00}, thereby validating our model in this 
respect. As for the pure DMTAP bilayer, Lewis et al. \cite{Lew01} 
have been able to extract the area per lipid in the low-temperature 
phase, finding $\langle A\, \rangle = 0.40$\,nm$^2$ at 25$^{\circ}$\,C. 
Studies of $\langle A\, \rangle$ above the main transition 
temperature are lacking, however.

\begin{figure}[tb]%
\includegraphics[width=6.0cm,bbllx=4,bblly=33,bburx=621,bbury=520,clip=]
{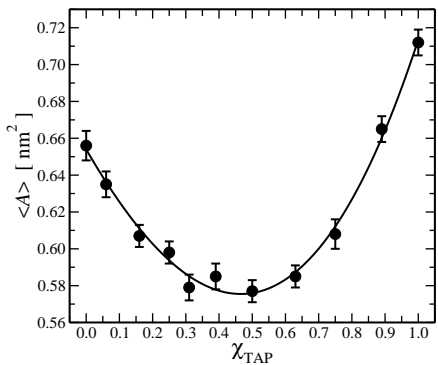} 
\caption{\label{fig.3}
Average area per lipid $\langle A \, \rangle $ as a function 
of the DMTAP mole fraction, $\chi_\mathrm{TAP}$.}
\end{figure}

We find that the average area per lipid shows a non-monotonic 
dependence on DMTAP mole fraction $\chi_\mathrm{TAP}$, with a pronounced 
minimum roughly at $\chi_\mathrm{TAP} = 0.5$, see Fig.~\ref{fig.3}. This 
behavior is not trivial, as modest amounts of the cationic 
DMTAP lead to a compression of the bilayer, while high concentrations 
lead to a major expansion of the membrane. More specifically 
we find that for $0 < \chi_\mathrm{TAP} \lesssim 0.8$, the average 
area per lipid is smaller than the corresponding counterpart 
for any of the pure lipid systems. Such a behavior cannot be 
explained by steric interaction alone but most likely is 
rather of electrostatic origin.

While no published experimental data exists (to the best of our 
knowledge) for the area per lipid of DMPC\,/\,DMTAP bilayer 
mixtures, similar behavior has been observed in related systems. 

For example, Zantl et al. \cite{Zan99b} considered DMPC\,/\,DMTAP 
monolayers using Langmuir-type film balance and found that, for 
pressures corresponding to the liquid-crystalline phase, the headgroup 
area decreased monotonically for small $\chi_\mathrm{TAP}$, then had a minimum 
at about the equimolar ratio ($\chi_\mathrm{TAP} \approx 0.5$), and 
increased for larger DMTAP mole fractions. This is in accord 
with our observations.

Another related work concerns Langmuir 
balance studies of mixed monolayers of zwitterionic 
palmitoyloleoylphosphatidylcholine (POPC) and cationic 
2,3-dimethoxy-1,4-bis($N$-hexadecyl-$N$;$N$-dimethyl-ammonium)butane 
dibromide (SS--1)~\cite{Saily01}. Although SS--1 is dicationic, 
one can qualitatively compare this system to our DMPC\,/\,DMTAP 
mixture. For the POPC\,/\,SS--1 system S{\"a}ily et al.~\cite{Saily01} 
found that the average area per lipid has a non-monotonic behavior 
with a minimum at $\chi_{SS-1} \approx 0.38$. They also found that 
this effect depends on the charge of the head group, and it 
disappeared when POPC (having a zwitterionic head group) was 
replaced by neutral dioleylglycerol (DOG).

Our results in Fig.~\ref{fig.3} suggest local extrema in 
$\langle A \, \rangle $ when $\chi_\mathrm{TAP}$ is between 0.16 and 0.5 
(in addition to global minimum at $\chi_\mathrm{TAP} = 0.5$). In addition 
to the above-mentioned POPC and SS--1 study~\cite{Saily01}, similar 
and even more dramatic effects have been observed for mixtures 
of POPC and sphingosine~\cite{Saily03}. The existence of critical 
concentrations in lipid membranes has also been theoretically 
postulated by Somerharju et al.~\cite{Virtanen:1998,Somerharju:1999}. 
In the present case, the local extrema for DMPC\,/\,DMTAP mixtures are 
within error bars, and therefore may be interpreted as fluctuations 
of $\langle A \, \rangle $. To study such features in more detail 
one needs to decrease the fluctuations in the average area per 
lipid by, e.g., increasing the system size~\cite{Lin00}. This, 
however, is beyond the scope of the present study.

Nevertheless, we decided to approach this issue from a different 
perspective. To determine the average area per lipid separately 
for the two different components, we used the Voronoi tessellation 
technique in two dimensions \cite{Patra03}. In Voronoi tessellation, 
we first calculated the center of mass (CM) positions for the lipids 
and projected them onto the $x$-$y$ plane. A point in the plane 
is then considered to belong to a particular Voronoi cell, if it 
is closer to the projected CM of the lipid molecule associated 
with that cell than to any other CM position. As there is no unique 
definition for the area per molecule in a multi-component system, 
it is clear that the Voronoi results should be considered as 
suggestive rather than quantitative, providing insight mainly 
of the trends.

\begin{figure}[tb]
\includegraphics[width=6.0cm,bbllx=4,bblly=33,bburx=621,bbury=520,clip=]
{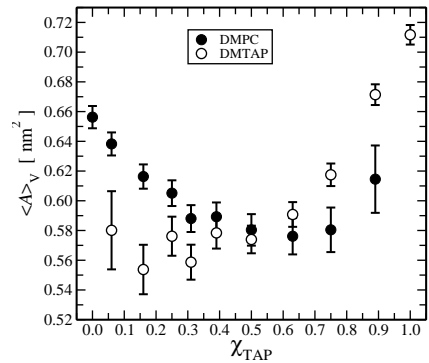}
\caption{\label{fig.4}
Average area per lipid $\langle A \, \rangle_{V} $ based on the 
Voronoi analysis in two dimensions. The results for DMPC and DMTAP 
are shown separately as a function of $\chi_\mathrm{TAP}$.}
\end{figure}

Figure~\ref{fig.4} demonstrates that the areas occupied 
by DMPC and DMTAP are distinctly different. For small $\chi_\mathrm{TAP}$, 
the area per DMPC is considerably larger than that of DMTAP. For 
larger DMTAP mole fractions above $\chi_\mathrm{TAP} = 0.5$, the situation 
is the opposite. This behavior is related to electrostatic effects 
and the ordering of acyl chains, and will be discussed in more 
detail in Sect.~\ref{sect:ordering_of_chains}. Here we only note 
that the fluctuations in $\langle A \, \rangle$ (see Fig.~\ref{fig.3}) 
at $0.1 \lesssim \chi_\mathrm{TAP} \lesssim 0.5$ arise from fluctuations 
in the area occupied by DMTAP. Whether this is a true result 
due to, e.g., clustering of lipids in this region remains to be 
resolved.

\subsection{Ordering of lipid acyl chains
            \label{sect:ordering_of_chains}}

Ordering of non-polar hydrocarbon chains in lipid bilayers is 
typically characterized by the deuterium order parameter $S_{CD}$ 
measured through $^2$H NMR experiments. If $\theta$ is the angle 
between a CD-bond and the bilayer normal, the order parameter is 
defined as 
\begin{equation}
\label{orderP}
S_{CD} = \frac{3}{2}\langle \cos^{2}\theta \rangle - \frac{1}{2} 
\end{equation}
separately for each hydrocarbon group. Since we employed 
a united-atom force field, the positions of the deuterium atoms 
are not directly available but have to be reconstructed from the 
coordinates of three successive nonpolar hydrocarbons, assuming 
an ideal tetrahedral geometry of the central CH$_2$ 
group~\cite{Chi95,Hof03}. In practice, we calculated 
$S_{CD}$ following the standard approach described 
elsewhere~\cite{Patra03}.

\begin{figure}[tb]
\includegraphics[width=6.0cm,bbllx=15,bblly=7,bburx=556,bbury=600,clip=]
{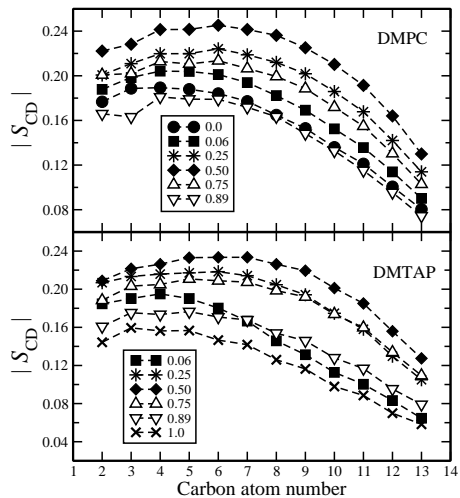} 
\caption{\label{fig.5}
Deuterium order parameter $|S_{CD}|$ averaged over $sn$--1 and 
$sn$--2 chains for DMPC (top) and for DMTAP (bottom) as a function 
of DMTAP mole fraction. Small carbon atom numbers correspond to 
those close to the headgroup.}
\end{figure}

Figure~\ref{fig.5} shows $|S_{CD}|$ averaged over the two 
similar atoms in the $sn$--1 and $sn$--2 chains, for both DMPC (top) 
and DMTAP (bottom) at different DMTAP concentrations. For the pure 
DMPC bilayer, we find $|S_{CD}| \approx 0.18$ close to the glycerol 
group of the molecule, in good agreement with recent experiments 
\cite{Pet00} and molecular dynamics simulation studies 
\cite{Pas01,Smo01}.

\begin{figure}[tb]
\includegraphics[width=6.0cm,bbllx=8,bblly=54,bburx=595,bbury=527,clip=]
{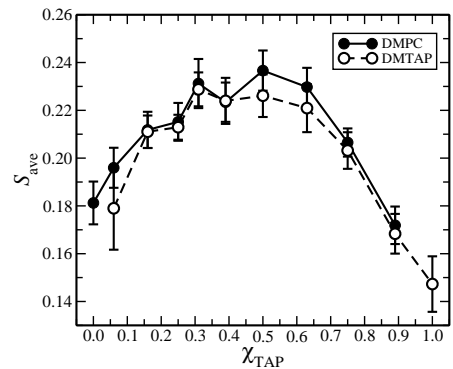} 
\caption{\label{fig.6}
Plateau order parameter $S_{\rm ave}$ calculated by averaging $S_{CD}$ 
over C2 to C8 hydrocarbons. Shown are $S_{\rm ave}$ for DMPC 
(solid lines with solid circles) and DMTAP (dashed lines with 
open circles).}
\end{figure}

The order parameter profile for the first seven hydrocarbons 
(from C2 to C8) is a kind of plateau, and the average value of 
$S_{CD}$ in this region, denoted by $S_{\rm ave}$, is shown
in Fig.~\ref{fig.6}. As expected, the results closely follow the 
change in the average area per lipid, i.e., a compression of the 
membrane is accompanied by enhanced ordering of nonpolar acyl chains. 
Results of similar nature have been observed, e.g., in bilayer 
mixtures of glycerophospholipids and cholesterol, where cholesterol 
both reduces the average area per molecule and enhances the 
ordering of lipid acyl chains \cite{Hof03}.

For pure DMPC the plateau value $S_{\rm ave}$ equals to
$0.181 \pm 0.009$, which is in very good agreement with the
experimentally measured value of $0.184$~\cite{Pet00}. For 
a pure DMTAP bilayer, the plateau value of $S_{CD}$ is considerably 
smaller, being about $0.147 \pm 0.011$. Therefore chains in 
a pure DMTAP bilayer are on average more disordered than in a DMPC 
system, in agreement with our findings for $\langle A \, \rangle$.

\subsection{Orientation of phosphatidylcholine headgroups
            \label{sect:orientation}}

Since the chemical structures of the acyl chains of DMPC 
and DMTAP are identical, it is obvious that the differences 
between their behavior are due to their headgroups. Since 
the headgroup of DMPC is zwitterionic (cf. Fig.~\ref{fig.1}), 
it possesses a dipole moment along the P\,--\,N vector. The 
electrostatic potential across a monolayer thus depends 
sensitively on the distribution of the angle $\alpha$ 
between the P\,--\,N vector and the interfacial normal $\vec{n}$ 
(where $\vec{n}$ has been chosen to point away from the bilayer 
center along the $z$ coordinate).

\begin{figure}[tb]
\includegraphics[width=6.0cm,bbllx=27,bblly=35,bburx=611,bbury=522,clip=]
{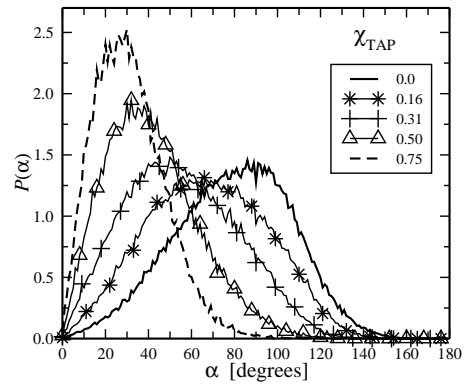} 
\caption{\label{fig.7}
Results for the probability distribution function $P(\alpha)$ vs 
the angle $\alpha$ between the P\,--\,N vector (of DMPC headgroups) 
and the bilayer normal.} 
\end{figure}

\begin{figure}[tb]
\includegraphics[width=6.0cm,bbllx=29,bblly=35,bburx=596,bbury=528,clip=]
{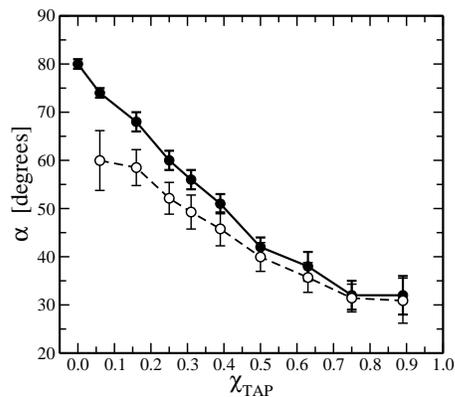} 
\caption{\label{fig.8}
The average angle $\langle \alpha \rangle $ between the P\,--\,N 
vector of DMPC and the bilayer normal, shown as a function of 
DMTAP mole fraction. Full circles: Results averaged over all DMPC 
molecules in a given system. Open circles: Results averaged over 
only those DMPC molecules that are beside DMTAP. 
See text for details. 
}
\end{figure}

Figure~\ref{fig.7} shows the probability distribution function 
$P(\alpha)$ for the angle in question. For a pure DMPC bilayer 
we find the distribution to be wide, thus allowing the P\,--\,N 
vector to fluctuate substantially, pointing at times in the 
direction of the membrane normal as well as into the bilayer 
interior. The average angle found in this case is about 
$(80 \pm 1)^{\circ}$ (see Fig.~\ref{fig.8}), i.\,e., the PC heads 
are on average almost parallel to the membrane surface. This is 
in agreement with experimental observations~\cite{Haus81,Scher89} 
as well as with recent computer simulations~\cite{Gab96,Smo99a,Pas99}.

Figure~\ref{fig.7} further reveals the major role played by DMTAP.
The addition of even a small amount of DMTAP leads to a pronounced 
re-orientation of the P\,--\,N dipole vector. As $\chi_\mathrm{TAP}$ is 
increased, the profile of the distribution becomes considerably 
narrower and its maximum shifts to smaller angles. This trend 
continues up to the high-concentration limit $\chi_\mathrm{TAP} \approx 0.75$, 
beyond which the distribution is essentially similar with the case 
found for $\chi_\mathrm{TAP} = 0.75$.

Results for the average angle $\langle \alpha \rangle$ between 
the P\,--\,N vector and the membrane normal shown in Fig.~\ref{fig.8} 
are consistent with this picture. On average, upon increasing 
$\chi _\mathrm{TAP}$, PC headgroups become more and more vertically 
oriented. Also this has been observed in several 
experiments~\cite{Scher89,Zan99b,Saily01,Saily03}. Moreover, 
our findings are in fairly good agreement with an atomistic MD 
study of a complex comprised of DNA and a mixture of DMPC and 
DMTAP~\cite{Ban99}, in which the average angle between the 
P\,--\,N dipole vector and the bilayer normal was found to be 
$(50 \pm 8)^{\circ}$ at an almost equimolar mixture of DMPC 
and DMTAP. In the present case without DNA, we found 
$\langle \alpha \rangle = (42 \pm 2)^{\circ}$ at $\chi_\mathrm{TAP} = 0.5$.

Perhaps surprisingly, the correlation between the average 
area per lipid (Fig.~\ref{fig.3}) and the re-orientation of 
the P\,--\,N dipole is not complete. As Fig.~\ref{fig.8} shows, 
the re-orientation extends by and large linearly up to 
$\chi_\mathrm{TAP} = 0.75$, while the membrane compression completes 
at $\chi_\mathrm{TAP} \approx 0.5$. This contradicts the conclusions of 
S{\"a}ily et al. \cite{Saily01} who studied POPC\,/\,SS--1 
cationic lipid mixtures using the Langmuir balance technique 
and suggested that the maximal average angle between the 
P\,--\,N dipole vector and membrane surface is achieved at the 
cationic lipid concentration that corresponds to the point where 
the membrane compression ends.

A closer inspection of Figs.~\ref{fig.3} and~\ref{fig.8} shows 
that the observed reduction of the average area per molecule is 
likely related to the re-orientation of PC dipoles, and this in 
turn is related to the role of electrostatic interactions between 
DMPC and DMTAP headgroups. To bridge the two issues, we propose 
the following schematic scenario. At small $\chi_\mathrm{TAP}$ where the DMTAP 
molecules are far apart and their mutual interaction is rather weak, 
we essentially suggest that the role of DMTAP is to re-orient
the headgroups of those DMPC molecules that are beside a DMTAP 
molecule. This favors more dense packing at small 
$\chi_\mathrm{TAP}$, leading to a reduction in $\langle A \, \rangle$,  
and consequently to a minimum in the area per molecule at 
intermediate concentrations as for large $\chi_\mathrm{TAP}$ the 
repulsive electrostatic interactions between TAP headgroups 
enforce $\langle A \,\rangle$ to be expanded.

To validate this scenario, we complemented our results in 
Fig.~\ref{fig.7} by calculating the probability distribution 
function $P(\alpha)$ for those DMPC molecules that are nearest 
neighbors to DMTAP.

As a criterion that a DMPC and a DMTAP form a pair,
we monitored the distance between PC phosphorus and TAP nitrogen.
For that, we first calculated the radial 
distribution functions (RDFs) between  pairs of P$_\mathrm{PC}$ and N$_\mathrm{TAP}$
and determined the distance $r_{nn}$ at which the RDF 
had its first minimum after the main peak (see also  
Sect.~\ref{sect:coordination_numbers}). 
The distance obtained in this fashion ($r_{nn} \approx 0.665$\,nm)
(and found not to depend on $\chi_\mathrm{TAP}$)
was applied to identify the DMPCs residing next to a DMTAP.
As shown in Fig.~\ref{fig.8},  the re-orientation of the 
P\,--\,N vector of these DMPCs
is considerably stronger at small DMTAP concentrations
as compared to that averaged over all DMPC lipids.

The above results imply that at small $\chi_\mathrm{TAP}$ \, the 
effect of DMTAP on the re-orientation is mainly local, i.e., the
alternating PC and TAP headgroups pack more tightly than 
in a pure DMPC system.
This idea is supported by the results for the radial distribution 
functions discussed in Sect.~\ref{sect:coordination_numbers}. Beyond 
the small-$\chi_\mathrm{TAP}$ regime, for intermediate concentrations 
$0.3 \lesssim \chi_\mathrm{TAP} \lesssim 0.5$, further increase in the 
concentration of DMTAP continues to increase the number of units 
composed of PC and TAP heads, thus favoring a reduction in 
$\langle A\,\rangle$. However, as repulsive electrostatic 
interactions between DMTAP molecules also become more and more 
important, the two effects compensate each other and 
$\langle A\,\rangle$ is found to be approximately constant. Finally, 
for large $\chi_\mathrm{TAP}$, the repulsive electrostatic interactions 
between TAP groups dictate the case discussed here and lead to 
an enhancement of the average area per molecule.

Though this picture does not account for the explicit influence 
of counter-ions, it grasps essence of the process. The effect 
of counter-ions is discussed separately in 
Sect.~\ref{sect:coordination_numbers}.

\subsection{Density profiles of lipid headgroups and chloride ions 
            \label{sect:density_profiles}}

To quantify the locations of charge groups and counter-ions, 
we computed the density profiles across the bilayer, separated 
into the different constituents of the system. The positions 
of all atoms in the system were determined with respect to the 
instantaneous center of mass position of the bilayer, exploiting 
mirror symmetry such that atoms with $z < 0$ were folded to 
$z > 0$ (the bilayer center being at $z = 0$).

\begin{figure}[tb]
\includegraphics[width=6.0cm,bbllx=36,bblly=2,bburx=386,bbury=603,clip=]
{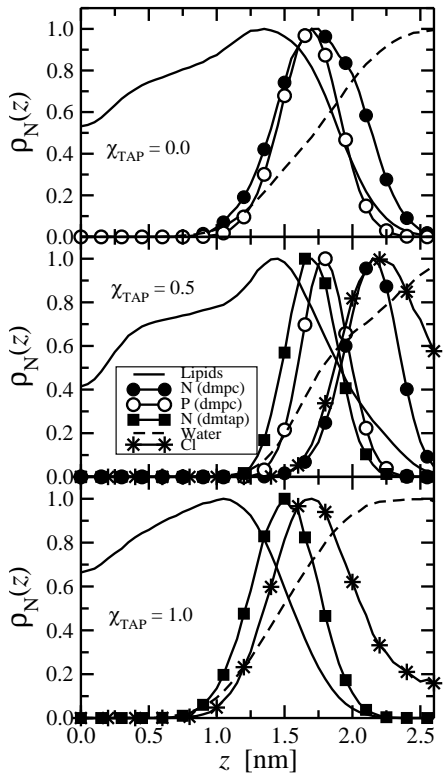} 
\caption{\label{fig.9}
Scaled number densities $\rho_N(z)$  for three DMPC\,/\,DMTAP 
mixtures with $\chi _\mathrm{TAP}=0.0$ (top), $\chi _\mathrm{TAP}=0.5$ (middle), 
and $\chi _\mathrm{TAP}=1.0$ (bottom). The case $z = 0$ corresponds to 
the center of the bilayer. 
}
\end{figure}

\begin{figure}[tb] 
\includegraphics[width=6.0cm,bbllx=29,bblly=35,bburx=596,bbury=528,clip=]
{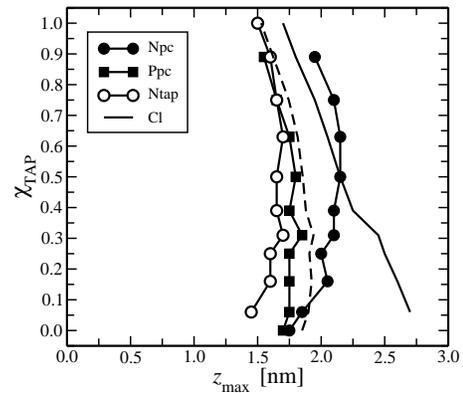} 
\caption{\label{fig.10}
Maxima of the density profiles, $z_{max}$, for phosphorus and 
nitrogen atoms from the DMPC headgroups, and of the density profile 
of chloride ions. The maxima are shown for the $z$ coordinate 
in the direction of the membrane normal, shown as a function of 
DMTAP molarity. The dashed line marks the position of the 
membrane-water interface determined from the condition that 
the densities of water and lipids (in Fig.~\ref{fig.9}) are 
equal.}
\end{figure}

Figure~\ref{fig.9} shows the scaled number densities $\rho_N(z)$ 
for a few selected cases. Additionally, we note that the essential 
information is given by the positions of the density maxima 
depicted in Fig.~\ref{fig.10}.

At small $\chi_\mathrm{TAP}$, the density maxima of the nitrogen 
and phosphorus atoms in the DMPC heads almost coincide 
(see Fig.~\ref{fig.9}). The density profile of nitrogen 
in DMPC is nevertheless broader and extends further out 
of the bilayer plane. For larger $\chi_\mathrm{TAP}$, the density 
profiles of phosphorus and nitrogen are distinctly separated, 
and nitrogen in particular extends rather deeply into the 
water phase. The TAP group represented by the nitrogen atom, 
however, is found to be deep in the bilayer. It seems 
obvious that these two issues are related, i.\,e., the density 
profiles of nitrogens in PC and TAP groups are well separated 
due to the electrostatic repulsion that essentially leads to 
the re-orientation of PC headgroups. These results are hence 
consistent with those in Sect.~\ref{sect:orientation} and 
reflect the dependence of DMPC headgroup orientation on 
$\chi_\mathrm{TAP}$.

Interestingly, while being attracted by the DMTAP headgroups, 
the chloride anions cannot penetrate the outer boundary of the 
bilayer formed by the DMPC choline groups. This is in a sense to be 
expected since the DMPC headgroup is longer than the TAP group 
and thus extends further outward from the bilayer. There is 
thus a significant amount of shielding of the chloride ions in 
the presence of DMPC. Only for an almost pure DMTAP layer are 
the chloride ions located in the vicinity of the TAP headgroups.

\subsection{Charge density, electrostatic potential, 
            and orientation of water}

\begin{figure}[tb]
\includegraphics[width=6.0cm,bbllx=40,bblly=2,bburx=441,bbury=604,clip=]
{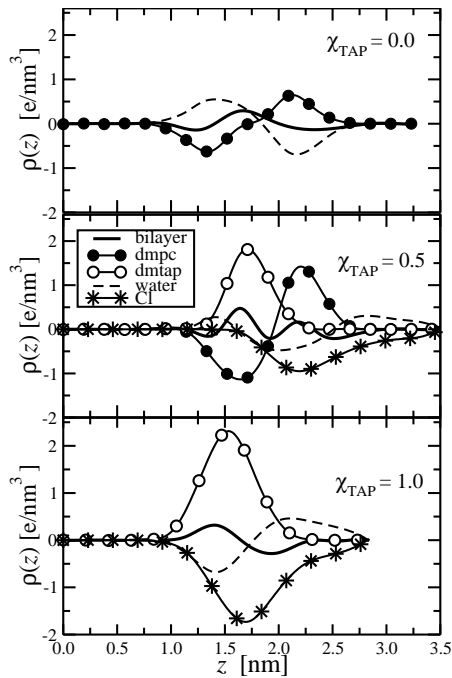} 
\caption{\label{fig.11}
Charge densities $\rho(z)$ across a single leaflet for 
$\chi_\mathrm{TAP}$ equal to 0.0 (top), 0.5 (middle), and 1.0 (bottom). 
The case $z = 0$ corresponds to the center of the bilayer. Charge 
densities are shown as solid lines. In addition, the component-wise 
contributions due to DMPC (full circle), DMTAP (open circle), 
water (dashed line), and chloride ions (star) are displayed. 
To reduce the noise in the data, the charge densities shown here 
were first fitted to splines~\protect\cite{Thi98}. The error bars 
are of the same size as the symbols. 
}
\end{figure}

The charge distribution shown in Fig.~\ref{fig.11} was calculated 
in the same fashion as the density profiles. The results are 
clearly reminiscent of the density profiles in Fig.~\ref{fig.9} 
and demonstrate the competition between charged PC and TAP 
groups, Cl anions, and water. The role of the TAP group is 
prominent.

\begin{figure}[tb]
\includegraphics[width=6.0cm,bbllx=20,bblly=39,bburx=597,bbury=521,clip=]
{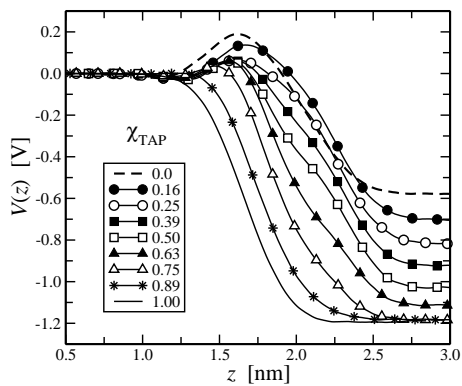} 
\caption{\label{fig.12}
Electrostatic potential $V(z)$ across cationic bilayers
at different DMTAP mole fractions.
}
\end{figure}

From the charge densities, we also computed the electrostatic 
potential across the bilayer. The results are shown in 
Fig.~\ref{fig.12} where the potential at the center of the
bilayer has been set to zero. For a pure DMPC bilayer we obtain 
$-0.578$\,V, which is in agreement with previous MD simulation 
studies~\cite{Chi95,Gab96,Smo99a}. Experimental data for 
phospholipid membranes ranges from $-0.2$\,V to $-0.6$\,V  
\cite{Hla73,Pic78,Fle86,Gaw92,Cla01}.

The charge of the DMTAP headgroup is mainly compensated by, 
depending on the value of $\chi_\mathrm{TAP}$, chloride ions or DMPC 
phosphate groups. Most of the electrostatic potential across 
the bilayer thus is not due to the DMTAP itself but rather 
due to the re-orientation of the DMPC headgroups. A clear 
indication of this is that the potential build-up saturates 
at $\chi_\mathrm{TAP} \simeq 0.75$, i.\,e., at the same value at which 
the distribution of headgroup orientation saturates 
(see Fig.~\ref{fig.8}). The total potential of the bilayer 
increases with increasing DMTAP concentration, with a difference 
of 0.6\,V between pure DMPC and pure DMTAP. This increase agrees 
well with the experimental data on cationic POPC\,/\,SS--1 
monolayers~\cite{Saily01}, and the authors offer the same 
explanation for their observations.

\begin{figure}[tb]
\includegraphics[width=6.0cm,bbllx=20,bblly=39,bburx=597,bbury=521,clip=]
{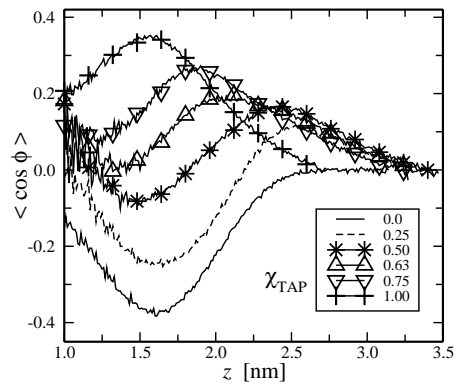} 
\caption{\label{fig.13}
Projection of the water dipole unit vector 
$\vec{\mu}(z)$ onto the interfacial normal $\vec{n}$, 
yielding $P(z) \equiv \langle \vec{\mu}(z)\cdot \vec{n} \rangle = 
             \langle \cos\phi \rangle$. Here $z = 0$ 
corresponds to the center of the bilayer and the bilayer 
normal $\vec{n}$ is chosen to point away from the bilayer 
center. 
}
\end{figure}

Many of the conclusions drawn from the charge density 
already follow from the number densities presented in 
Sect.~\ref{sect:density_profiles}, since for charged particles 
number density and charge density are trivially related. Water, 
however, has an additional internal degree of freedom, and 
a quick discussion of the orientation of the water molecules 
seems appropriate. As seen from Fig.~\ref{fig.13}, the average 
direction of the water dipoles in the membrane-water interface 
region is inverted for $\chi_\mathrm{TAP} = 0.5 \to \chi_\mathrm{TAP} = 1.0$. 
This is closely related with the familiar ``hump'' close to the 
interface, which is due to a subtle imbalance between the 
orientation of the water molecules and lipid headgroups~\cite{Chi95}. 
At higher DMTAP concentrations this ``hump'' disappears. A related 
issue concerns the pure DMTAP bilayer, in which case the density 
profile of water penetrates rather deep into the membrane 
(see Fig.~\ref{fig.9}), extending up to the interface 
region between the polar TAP group and the hydrophobic core. 
This is in accord with the interpretation of Fourier-transform 
infrared spectroscopic measurements by Lewis et al. 
\cite{Lew01}.

\subsection{Radial distribution functions and coordination numbers 
            \label{sect:coordination_numbers}}

To characterize the structure of the membrane-water interface 
region in more detail, we computed various radial distribution 
functions (RDFs) between the atoms in the headgroups, Cl, and 
the oxygens. Here we briefly discuss the most relevant results.

The RDFs between the center of mass positions indicated that the leading 
(main) peak for DMPC\,--\,DMPC and DMTAP\,--\,DMTAP pairs was 
rather broad and at about 1.0\,nm (data not shown). For 
DMPC\,--\,DMTAP pairs, however, the main peak of the RDF was 
much closer, at around 0.7\,nm. This supports the conclusion 
made in Sect.~\ref{sect:orientation}, i.e., DMPC and DMTAP form 
units that allow more dense packing than in a pure DMPC bilayer.

The N$_{\rm PC}$\,--\,N$_{\rm PC}$ and N$_{\rm TAP}$\,--\,N$_{\rm TAP}$ 
pairs were found to be rather far apart, the position of their main 
peak being at about 0.83\,nm, while the N$_{\rm PC}$\,--\,N$_{\rm TAP}$ 
pair was slightly closer (0.8\,nm). The positions of the main peaks 
did not depend on $\chi_\mathrm{TAP}$. As for the RDFs of the phosphorus 
atoms in the PC headgroups, its main peak with respect to N$_{\rm PC}$ 
and N$_{\rm TAP}$ was found to be at a much closer distance, at 
0.465\,nm for N$_{\rm PC}$ and 0.485\,nm for N$_{\rm TAP}$. Again, 
the positions of these peaks did not depend on the DMTAP 
concentration.

\begin{figure}[tb]
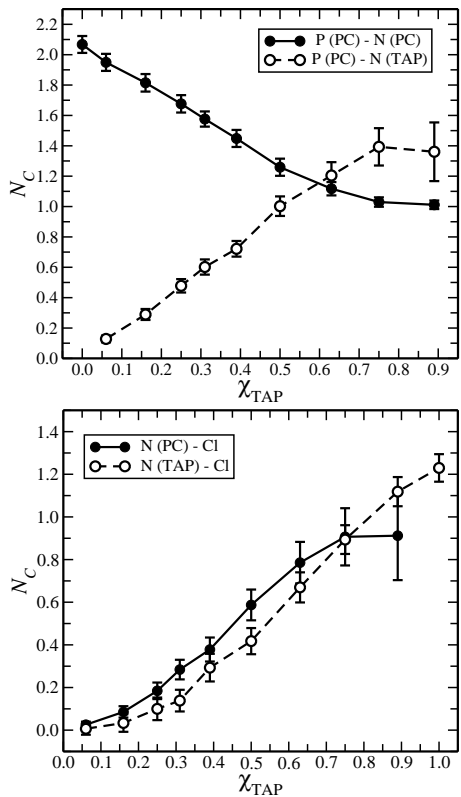

\includegraphics[width=6.0cm,bbllx=32,bblly=36,bburx=582,bbury=522,clip=]{coord_Ppc_N_sum.eps} 
\includegraphics[width=6.0cm,bbllx=30,bblly=38,bburx=588,bbury=522,clip=]{coord_N-Cl_sum.eps} 
\caption{\label{fig.14}
Top: Coordination numbers $N_C$ of DMPC phosphorus with DMPC 
nitrogen (solid line with solid circles) and with DMTAP nitrogen 
(dashed line with open circles) plotted versus $\chi_\mathrm{TAP}$.
Bottom: Coordination numbers $N_C$ of DMPC nitrogen 
(solid line with solid circles) and DMTAP nitrogen 
(dashed line with open circles) with Cl ions.
}
\end{figure}

We also calculated the  coordination numbers for the phosphorus and 
nitrogen atoms at different DMTAP concentrations. These are shown in
Fig.~\ref{fig.14} (top). It turns out that in the range from 
$\chi _\mathrm{TAP} = 0$ to 0.5 the PC nitrogens are to an increasing 
extent being replaced by N$_{\rm TAP}$ in the vicinity of P. 
This has twofold consequences: First, the electrostatic attraction 
between N$^+$ (TAP) and P$^-$ (PC) enhances the compression 
of the bilayer for $0.0 < \chi_\mathrm{TAP} \lesssim 0.5$ 
(see Fig.~\ref{fig.3}). Second, the decreasing coordination 
number for P\,--\,N$_{\rm PC}$ with $\chi_\mathrm{TAP}$ support the 
view that the DMPC nitrogens are pushed towards water, thereby 
PC headgroups are re-oriented to a more vertical alignment 
with respect to the membrane plane (Fig.~\ref{fig.8}).

We conclude this section with a discussion of the location 
of chloride counter-ions. The positions of the main peaks of 
the RDFs for both types of nitrogens (in PC and TAP groups)
with respect to Cl ions are identical, about 0.475\,nm, and 
do not depend on $\chi_\mathrm{TAP}$ (data not shown). 
In Fig.~\ref{fig.14} (bottom) we plot the coordination numbers 
of chlorides in the vicinity of both types of nitrogens as 
a function of DMTAP concentration. 
The figure confirms that Cl ions 
are preferentially bound to PC nitrogens rather than to N$_{\rm TAP}$. 
This holds up to a DMTAP mole fraction of about 0.75. 
The explanation for this is straightforward: As $\chi_\mathrm{TAP}$ 
increases, the PC headgroups become more and more vertically 
oriented with respect to the bilayer plane. This, in turn, makes 
PC nitrogens more easily accessible for the Cl ions. In contrast, 
small TAP heads are located much deeper in the membrane surface 
region than the PC heads and therefore are able to attract fewer 
chlorides regardless of the fact that TAP heads carry a net 
positive charge. Interestingly, when the re-orientation of 
PC heads is accomplished (at $\chi_\mathrm{TAP} \approx 0.75$) the 
coordination number for N$_{\rm PC}$\,--\,Cl pairs seems to 
saturate, see Fig.~\ref{fig.14} (bottom).

\section{Summary and conclusions}

Drug delivery and gene therapy have attracted substantial 
interest due to their importance in treating human diseases. 
As far as experimental work is concerned, particular attention 
has been paid to non-viral delivery vectors such as cationic 
liposomes characterized by a number of desired properties 
such as high efficiency and lack of toxicity. 
Consequently, it is surprising how little is known about the 
atomic-level details of cationic lipid bilayers. Essentially, 
this is due to the lack of molecular simulations of these 
systems, the study by Bandyopadhyay et al.~\cite{Ban99} 
being the only exception, to the authors' knowledge, in 
this regard.

As a first step toward a detailed understanding of cationic 
membrane--DNA complexes on atomic level, we have employed 
extensive molecular dynamics simulations of lipid bilayer 
mixtures composed of cationic DMTAP and neutral (zwitterionic) 
DMPC. Such binary DMPC\,/\,DMTAP  mixtures have been studied 
widely through experiments, and have been shown to form stable 
complexes with DNA~\cite{Art98,Zan99a,Zan99b,Poh00}. In the 
present work, we have focused on the influence of the composition 
of the cationic bilayer on its structural and electrostatic 
properties. For this purpose we studied numerous DMPC\,/\,DMTAP 
mixtures in the liquid-crystalline phase by varying the mole 
fraction of DMTAP, $\chi_\mathrm{TAP}$, from the pure DMPC to the pure 
DMTAP bilayer.

We have found that the properties of the DMPC\,/\,DMTAP bilayer 
mixture are largely dominated by the electrostatic properties of 
the headgroup region around the membrane-water interface. Most 
notably, our results indicate that there is a strong interplay 
between the PC and TAP groups together with the Cl counter-ions 
that concentrate in the vicinity of the bilayer-water interface. 

The interplay between the PC and TAP groups leads to a number 
of intriguing observations. The key factor here is the 
re-orientation of PC groups due to an introduction of DMTAP 
in the bilayer. The re-orientation of the PC headgroups 
arises from electrostatic interactions that lead phosphate and 
choline groups to rearrange their positions with respect to 
the cationic TAP. This effect is enhanced as $\chi_\mathrm{TAP}$ is 
increased, and extends up to large molar fractions of approximately
$\chi_\mathrm{TAP}= 0.75 $. Beyond this limit a further increase 
of DMTAP concentration has no additional effect on the 
orientation of PC headgroups. Interestingly, at small $\chi_\mathrm{TAP}$ 
the effect of the re-orientation is of  local nature,
i.e., the P\,--\,N dipoles of DMPCs beside DMTAP molecules  
re-orient considerably.

\begin{figure}[tb]
\includegraphics[width=6.0cm,bbllx=0,bblly=0,bburx=620,bbury=582,clip=]
{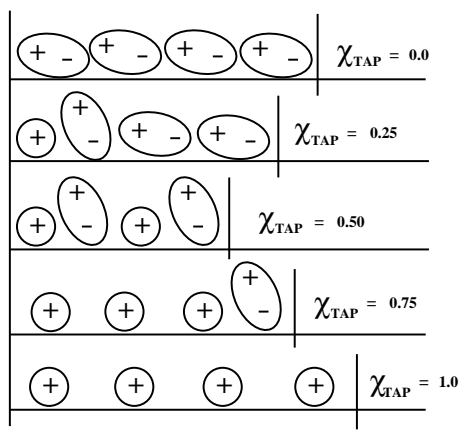} 
\caption{\label{fig.15}
A proposed schematic picture of the observed change in the 
area per lipid vs $\chi_\mathrm{TAP}$. Only headgroups of the lipids 
are shown here, and water (not shown) is above the headgroups. 
}
\end{figure}

At small molar fractions of DMTAP, the re-orientation of 
PC dipoles leads to considerable compression of the bilayer 
as alternating PC and TAP groups are able to pack more tightly 
than in a pure DMPC bilayer. The minimum of the area per lipid 
at $\chi_\mathrm{TAP} \approx 0.5$ is about 12\,\% smaller than in 
the pure DMPC bilayer. A further increase of $\chi_\mathrm{TAP}$ 
leads to major expansion of the bilayer. This is essentially 
due to an increasing number of neighboring TAP groups whose 
cationic nature leads to repulsive electrostatic interactions 
that do not favor close packing. As expected, the ordering of 
acyl chains closely follows the change in the area per lipid. 
When these results are summarized, the present view of the 
average area per lipid coupled to the re-orientation of the 
headgroups can be summarized schematically as in Fig.~\ref{fig.15}.

In view of future studies of DNA-membrane systems, it is important
to pay attention to the influence of DMTAP on the electrostatic 
properties of the membrane, including the increase in the 
electrostatic potential across the bilayer and the ordering of 
water in the vicinity of the membrane-water interface. 
Another often ignored  aspect of electrostatics is that the 
ionic buffer liquids may affect membranes
significantly~\cite{Boeckmann:2003}.

Perhaps the most significant observation in this study is the spatial 
re-arrangement of PC and TAP headgroups which is expected to play 
a significant role in the condensation of DNA onto the membrane 
surface. The cationic TAP and choline groups then play a key 
role as the anionic phosphate groups of DNA come into contact 
with the membrane. While the present study clarifies some of 
the underlying questions related to binary mixtures of cationic 
and neutral (zwitterionic) lipid membranes, further atomic-level 
studies are essential to resolve other important issues related to 
DNA-membrane systems, such as the influence of salt and its 
screening effects, and the stability and interface properties 
under those conditions. Work in this direction is under way.

\section{Acknowledgments}

We wish to thank M.\,T. Hyv{\"o}nen, P. Niemel{\"a}, 
and X. Periole for fruitful discussions. This work has 
been supported by the Academy of Finland Grant Nos. 
202598 (A.\,G.), 54113 (M.\,K.), and 80246 (I.\,V.), 
the Academy of Finland through its Center of Excellence 
Program (A.\,G., I.\,V.), and by the European Union through 
the Marie Curie fellowship HPMF--CT--2002--01794 (M.\,P.).
We would also like to thank the Finnish IT Center for Science (CSC)
and the HorseShoe (DCSC) supercluster computing facility at 
the University of Southern Denmark for computer resources.


\end{document}